# Local magnetic inhomogeneities observed via $^{75}$As NMR in Ba(Fe$_{1-x}$Ni$_x$)$_2$As$_2$ with $H_0$ ⊥ c-axis

A P Dioguardi[1], N apRoberts-Warren[1], A C Shockley[1], S L Bud'ko[2], N Ni[2], P C Canfield[2], and N J Curro[1]

[1]Department of Physics, University of California Davis
One Shields Ave., Davis, CA 95616

[2]Ames Laboratory & Iowa State University
Ames, IA 50011

E-mail: dioguardi@ms.physics.ucdavis.edu

**Abstract.** We present field-swept $^{75}$As Nuclear Magnetic Resonance (NMR) measurements in the spin-density wave (SDW) state of Ba(Fe$_{1-x}$Ni$_x$)$_2$As$_2$ for x=0.0072. The large single crystals were aligned with the external field $H_0$ perpendicular to the c axis. The spectra are increasingly broadened as a function of doping, and are well fit by a model of a commensurate SDW with local impurities.

## 1. Introduction

The discovery of high temperature superconductivity by Bednorz and Müller in 1986[1] raised both hopes and questions about the nature of the Cooper pairing mechanism. The inability of phonon pairing to account for such high transition temperatures as well as the proximity of superconductivity to antiferromagnetism lead many physicists to believe that magnetism might be the key to understanding the high $T_C$ in these new materials.[2] However, to this day there is still no generally accepted microscopic theory of high temperature superconductivity.

Twenty two years later in 2008 the 1111 iron arsenide family of high temperature superconductors was discovered by H. Hosono's group.[3] Soon thereafter, the discovery of the subclass known as the 122 iron arsenide materials by M. Rotter et al.[4] provided researchers with the opportunity to study large single crystals grown by flux methods. The Fe-As high-temperature superconductors provide researchers with a new system in which superconductivity is found to border on and even possibly coexist[5][6] with a long range antiferromagnetic order of iron moments.

NMR has been shown to be a powerful technique to investigate strongly correlated electron systems such as the ferropnictides. Since NMR is sensitive to both local magnetic fields—produced by SDW ordering of the Fe-As superconductors—and the local electronic structure—via the nuclear quadrupolar coupling to the electric field gradient (EFG)—the technique remains central to the study of Fe-As high temperature superconductors. Recently, $^{75}$As NMR has been used to study the effect of

Co and Ni dopants in the 122 system.[7][8] The broadening of the $^{75}$As spectra has been interpreted both as (a) the presence of incommensurate AF order, and (b) commensurate AF order with local perturbations in the hyperfine field surrounding the dopants. Here we present spectra of the $^{75}$As for $\mathbf{H_0} \perp$ c-axis which support the latter interpretation.

## 2. Nuclear Magnetic Resonance Results

Large single crystals of Ba(Fe$_{1-x}$Ni$_x$)$_2$As$_2$ were grown in Fe-As flux according to [6], [9], and [10]. The plate-like crystals were aligned with the external field perpendicular to the c-axis of the crystal inside of the NMR coil. Field swept spectra were obtained at T = 20K and $f$ = 48.28 MHz by integrating the area under the spin echo as a function of applied field utilizing the same apparatus as [8]. $^{75}$As has nuclear spin I=3/2 and the resonance condition is governed by the nuclear spin Hamiltonian:

$$\mathrm{H} = \gamma\hbar\hat{\vec{I}} \cdot \vec{H}_0 + \frac{h\nu_c}{6}\left[3\hat{I}_c^2 - \hat{I} - \eta\left(\hat{I}_a^2 - \hat{I}_b^2\right)\right] + \mathrm{H}_{hf} \quad (1),$$

where $\gamma$ = 0.7292 kHz/G is the gyromagnetic ratio; $H_0$ is the applied field; $\hat{I}_a$ are the nuclear spin operators with respect to the principal axes of the EFG; $\nu_c = 3eQV_{cc}/2I(2I-1)h \cong 2.5$MHz, with e the proton charge, $Q$ the nuclear quadrupole moment, $V_{cc}$ the component of the EFG ∥ c-axis, $I$ the nuclear spin, and $h$ planck's constant; $\eta$ is the asymmetry parameter of the EFG tensor and H$_{hf}$ is the hyperfine Hamiltonian, given by

$$\mathrm{H}_{hf} = \gamma\hbar\hat{\vec{I}} \cdot \sum_{i \in nn} \vec{\vec{B}}_i \cdot \vec{S}(\vec{r}_i) \quad (2),$$

where $\mathbf{B}_i$ is the hyperfine coupling tensor and $\mathbf{S}(r_i)$ is summed over the four nearest neighbor Fe spins.

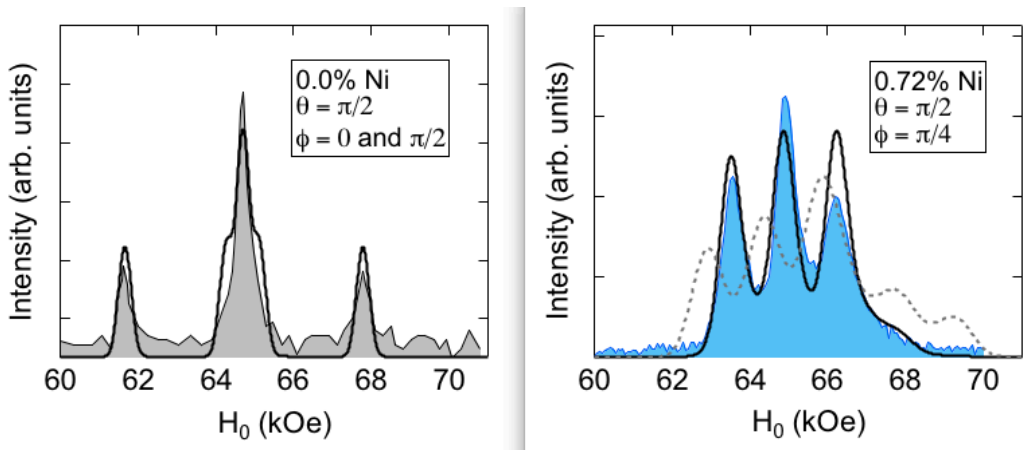

Figure 1. (Color Online) Field-swept spectra of Ba(Fe$_{1-x}$Ni$_x$)$_2$As$_2$ with H$_0$ ⊥ c-axis for x = 0 (grey, left) and x = 0.0072 (blue, right). Angles θ and φ refer to the crystal orientation with respect to $\mathbf{H_0}$ used to calculated the computer simulations (solid black lines). Grey dashed lines refer to incommensurate calculations. The x = 0 data is reproduced from [11].

Below the temperatures $T_S(x)$ and $T_N(x)$, Ba(Fe$_{1-x}$Ni$_x$)$_2$As$_2$ exists in the orthorhombic *Fmmm* crystal structure and a SDW magnetically ordered state respectively.[9][10] In this magnetically ordered state the internal hyperfine field modifies the resonant conditions via $H_{hf}$. The observed field-swept spectra are denoted by the filled colored lines in Figure 1. The first plot with x = 0 is reproduced from Kitagawa et al.[11] for purposes of comparison to the parent compound. The spectra broaden rapidly as function of doping and it is clear that the crystals become twinned below $T_S(x)$ as described in [12].

## 3. Computer Simulations

To better understand the subtleties of these field-swept NMR spectra, computer simulations were carried out to calculate resonant field histograms for both the incommensurate and local impurity models. The hyperfine coupling tensor elements were drawn from [11] and combined with the commensurate ordering wave vector Q = ($\pi$/a,0,0)[13] to calculate the internal field $H_{int}$ at the $^{75}$As site. The magnetic order was then modified by Gaussian suppression around randomly placed impurities of appropriate numbers on a 100 x 100 lattice utilizing the same parameters as [8]. The difference in the resulting magnetic order as well as the in-plane components of the internal hyperfine field near a Nickel dopant are represented in Figure 2.

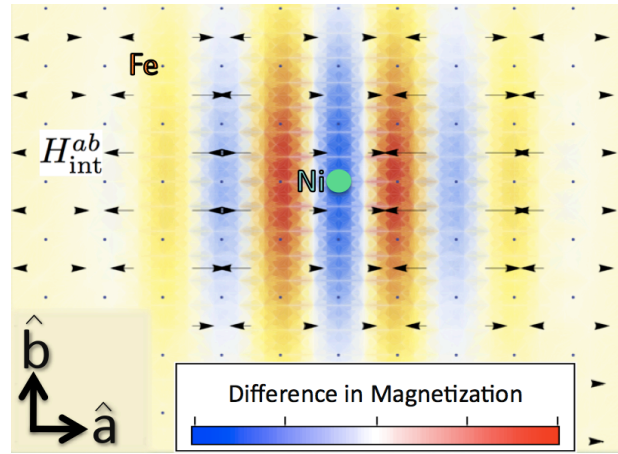

Figure 2. (Color online) The local perturbation to the SDW order surrounding an impurity is represented by a continuous Blue-to-Red color scale. The in-plane components of the hyperfine field are denoted by black arrows. The dopant influences the magnetic order out to multiple lattice constants.

After calculating the internal field as a function of position, we determine the resonance field by solving Eq. 1 using first order perturbation theory:

$$\gamma \left| \vec{H}_0 + \vec{H}_{int} \right| = f - p \frac{v_c}{2} \left( 3\cos^2\theta - 1 + \eta \sin^2\theta \cos(2\phi) \right), \quad (3)$$

where $\theta$ and $\phi$ are the angles with respect to the total field $\boldsymbol{H}_{tot} = \boldsymbol{H}_0 + \boldsymbol{H}_{int}$. The resulting histograms are represented by the black curves in Figure 1. It is clear that the internal hyperfine field begins to be canted into the plane around a dopant atom and only points along the [100] (a-axis). As a result the orientation of the crystal with respect to the external field will modify the field swept spectra due to both the Zeeman interaction and the quadrupolar interaction. We also show the simulated spectrum

(dashed grey lines in Figure 1.) for an incommensurate magnetization **S(r)** as described in [7]. Clearly, the local perturbation picture fits the data best.

## 4. Conclusion

In summation, the $^{75}$As NMR field swept spectrum was observed in the SDW state of Ba(Fe$_{1-x}$Ni$_x$)$_2$As$_2$ for x = 0.0072 and compared to the parent compound[11] with the external field applied along the [110] direction and the [100]+[010] respectively. Crystal twinning results in this combination of applied field directions.[12] The spectra broaden rapidly as a function of Nickel substitution as the internal hyperfine field is canted along the [100] direction. Computer simulations within an impurity model based on Gaussian suppression of the local moments are able to reproduce the field swept spectra. The simulations are sensitive to the applied field direction with respect to the magnetic order, and therefore, the crystal axes.

**Acknowledgements**:

The authors thank the Yukawa Institute for Theoretical Physics at Kyoto University. Discussions during the YITP workshop YITP-W-10-12 on "International and Interdisciplinary Workshop on Novel Phenomena in Integrated Complex Sciences: from Non-living to Living Systems" were useful to complete this work. This workshop was supported in part by the Grant-in-Aid for the Global COE Program "The Next Generation of Physics, Spun from Universality and Emergence" from the Ministry of Education, Culture, Sports, Science and Technology (MEXT) of Japan.
    This work was supported by the U.S. Department of Energy, Office of Basic Energy Science, and Division of Materials Sciences and Engineering. The research was performed at Ames Laboratory. Ames Laboratory is operated for the U.S. Department of Energy by Iowa State University under Contract No. DE-AC02-07CH11358.
    The authors would also like to thank the conference organizers as well as ICAM/I2CAM for travel funding. A. P. Dioguardi would especially like to thank his advisor and mentor Nicholas J. Curro for sharing his knowledge of NMR as well as his contagious excitement for learning.